\documentstyle[preprint,flushrt]{aastex}
\begin{document}
\newcommand{\redshift} {\left( \frac{1+z}{10} \right)}
\newcommand{\SFR} {{\rm \left( \frac{SFR}{1 M_{\odot} \, yr^{-1}}
\right)}}  
\newcommand{\Bfield} {{\rm \left( \frac{B}{10 \mu G} \right) }}
\newcommand{\efficiency} {\left( \frac{\epsilon}{0.1} \right)}
\newcommand{\freq} { \left( \frac{\nu}{1 \, {\rm GHz}} \right)}
\newcommand{\radrel} { {\rm min} \left[ \redshift^{-4},
\left( \frac{U_{\gamma}}{ 4.2 \times 10^{-9} \, {\rm erg \, cm^{-3}}}
\right)^{-1} \right] }
\newcommand{\density} {\left( \frac{n_{g}}{1 \, {\rm cm^{-3}}}
\right)}
\newcommand{\sfryr}{{\rm M_{\odot} \, yr^{-1}}}

\title{Probing High Redshift Radiation Fields with
Gamma-Ray Absorption} 
\author{S. Peng Oh\altaffilmark{1}\\
Princeton University Observatory, Princeton, NJ 08544; peng@astro.princeton.edu}
\altaffiltext{1}{Current address: Theoretical Astrophysics, Mail Code
130-33, California Institute of Technology, Pasadena, CA 91125; peng@tapir.caltech.edu}
\begin{abstract}
The next generation of gamma-ray telescopes may be able to observe
gamma-ray blazars at high redshift, possibly out to the epoch of
reionization. The spectrum of such sources should exhibit an
absorption edge due to pair-production against UV photons along the
line of sight. One expects a sharp drop in the number density of UV photons at
the Lyman edge $\epsilon_{L}$. This implies that the universe becomes
transparent after gamma-ray photons redshift below $E \sim
(m_{e} c^{2})^{2}/\epsilon_{L} \sim 18 {\rm GeV}$. Thus, there is only a limited redshift interval over which GeV
photons can pair produce. This implies that any observed absorption
will probe radiation fields in the very early universe, regardless of
the subsequent star formation history of the universe. Furthermore,
measurements of differential absorption between blazars at
different redshifts can cleanly isolate the opacity due to UV emissivity at high
redshift. An observable absorption
edge should be present for most reasonable radiation fields with
sufficient energy to reionize the universe. Ly$\alpha$ photons may
provide an important component of the pair-production opacity. Observations of a number of blazars at
different redshifts will thus allow us to probe the rise in comoving
UV emissivity with time.  
\end{abstract}

\section{Introduction}

Our knowledge of UV radiation fields and energy injection into the IGM
at $z>5$ is fairly tenuous. There are two main constraints: observations of the integrated background light (Madau
\& Pozzetti 2000, Bernstein et al 1999), and the fact that no
Gunn-Peterson trough is observed in the spectra of the
highest-redshift quasar to date (Fan et al 2000), implying that the
universe must be reionized by $ z = 5.8$. The upcoming Next Generation Space Telescope
(NGST) will be able to image high-redshift star clusters or AGNs in rest frame
UV continuum emission (Haiman \& Loeb 1997,1998), and their redshifts may
be obtained via H$\alpha$ observations (Oh 1999). Nonetheless, the
redshift-binned number counts will be fairly sparse, and one is
unlikely to probe sufficiently far down the luminosity function
to get a good measure of the comoving emissivity as a function of
redshift. 

Observations of gamma-ray blazars (``grazars'') probe extragalactic IR and UV
radiation fields, by observing the pair production opacity to $\gamma$
rays at the high energy end (Gould \& Schreder 1967, Stecker, De Jager \& Salamon 1992, Madau
\& Phinney 1996, Primack et al 1999). All theoretical models have confined
their predictions to low redshift grazars, with the exception of
Salamon \& Stecker (1998), who computed the $\gamma$-ray opacity up to
z=3. They concluded that because the stellar emissivity peaks between
z=1 and z=2, the $\gamma$-ray opacity shows little increase at high
redshift, and thus is not dependent on the initial epoch of galaxy
formation.

To date, EGRET has detected 66 gamma-ray loud blazars
(Hartman et al 1999), out to redshifts $z>2$. The next generation of gamma-ray telescopes (GLAST, CELESTE,
STACEE, MAGIC, HESS, VERITAS, and Milagro) should greatly enlarge this
sample. If the low redshift correlation between black hole mass and
bulge mass (Magorrian et al 1998) continues to high redshift, then it is
possible that high-redshift halos could host mini-quasars (Haiman \&
Loeb 1998, Haehnelt, Natarajan \& Rees 1998), which should be
detectable in rest frame UV emission by NGST and X-ray emission by
Chandra (Haiman \& Loeb 1998, 1999) in the redshift range
$z \sim 5-15$. This raises the exciting possibility that grazars
could be detected at similarly high redshifts. It is worth noting that EGRET has detected $\sim 56$ sources at high
Galactic latitudes $b > 10^{\circ}$ (Mukherjee, Grenier \& Thompson
1997), with no known counterparts at other wavelengths. Their spatial
distribution and log N- log S plot can be well fit by a Galactic
component plus an isotropic, extragalactic contribution. Some of
these may well be unidentified high-redshift blazars.   
   
In this paper, I point out that if grazars are detected at high
redshifts $z>3$, the pair production opacity to gamma ray photons can be used
to probe the comoving emissivity longward of the Lyman break at these
extremely high redshifts, independent of the star formation rate at
lower redshifts. Due to the small escape fraction of ionizing photons $f_{esc} < 5 \%$ from host
galaxies, as well as the high photoelectric opacity of the IGM at
these wavelengths, the comoving number density of UV photons exhibits
a sharp drop at the Lyman edge at all redshifts. Thus, there is only a limited
pathlength over which a gamma-ray photon can pair produce against UV photons,
before it redshifts to energies which require UV photons above the
Lyman edge for pair production to take place. For $z < z_{break}$, the
universe becomes optically thin to the gamma-ray photon. Thus, the detection of an absorption
edge in a high-redshift grazar places an immediate constraint on the
mean radiation field over the redshifts $z_{break} < z <
z_{s}$. Furthermore, measurement of the different absorption at a
given observed energy between blazars at redshifts $z_{1},z_{2}$
places an immediate constraint on the radiation field in the redshift
range $z_{1} < z < z_{2}$. Detection of grazars at a number of redshifts would then
enable one to probe the UV emissivity history of the universe.   

In all numerical estimates, we assume a background
cosmology given by the 'concordance' values of Ostriker \& Steinhardt
(1995): $(\Omega_{m},\Omega_{\Lambda},\Omega_{b},h,\sigma_{8
h^{-1}},n)=(0.35,0.65,0.04,0.65,0.87,0.96)$.

\section{Gamma-Ray Blazars}

A detailed study of the detectability of high-redshift blazars is
beyond the scope of this paper. In this section, I merely show that
it is plausible that GLAST will be able to detect high redshift blazars. 

With a point source sensitivity of $S(E > 100 {\rm MeV}) \sim 2 \times
10^{-7}$photons ${\rm s^{-1} cm^{-2}}$, EGRET has detected $\sim 66$
high-redshift blazars out to $z > 2$ (Hartman et al 1999). The
associated gamma-ray luminosities correspond to $L_{\gamma} = 10^{46}-10^{49}
\, {\rm erg \, s^{-1}}$, and typically dominate the bolometric
luminosity of the AGN, with $L_{\gamma}/L_{B} \sim 1-1000$. The
upcoming gamma-ray telescope GLAST (see http://glast.gsfc.nasa.gov)
will be 2 orders of magnitude more sensitive, with a detection
threshold of $S(E > 100 {\rm MeV}) \approx 2 \times 10^{-9}$photons ${\rm s^{-1}
cm^{-2}}$ for a 5 $\sigma$ detection with a 50 hour integration, and
$S(E > 1 {\rm GeV}) \approx  10^{-10}$ photons ${\rm s^{-1} cm^{-2}}$ (these
thresholds correspond to the same detection limit for a $L_{\nu}
\propto \nu^{-\alpha}$ source spectrum where $\alpha=1$). Goals for
GLAST include a broad energy
coverage from $10 \, {\rm MeV}- > 300 \, {\rm GeV}$, with a spectral resolution of $ \sim 2 \%$  in the $>$ 10
GeV range, a field of view of $>$ 3 sr, and a source location
determination accuracy of 30 arcsec- 5 arcmin.  During its lifetime,
it will perform an all-sky survey similar to that conducted by EGRET. Sources of the same or somewhat fainter
luminosity as those detected by EGRET may be seen by GLAST out to
extremely high redshifts, $z \sim 10$. 

Will such luminous sources
will be present at high redshift? If the AGN is assumed to emit all
its energy at gamma-ray wavelengths at the Eddington luminosity, the
inferred black hole mass is extremely high, $M_{bh} = 10^{10} M_{\odot} (L_{edd}/10^{48} {\rm erg s^{-1}})$. However,
there are two reasons why gamma ray sources of high apparent luminosity do not require such massive black holes: (i)
even if all photons are radiated isotropically, if most of the
radiation emerges at high energies (as appears to be the case in gamma-ray
blazars), then Klein-Nishina effects must be taken into account
(Dermer \& Gehrels 1995). The inferred black hole mass, given by $
M_{8}^{KN} \ge \frac{3 \pi d_{L}^{2} (m_{e} c^{2})} {2 \cdot 1.26 \times
10^{46} {\rm erg \, s^{-1}} } \frac{F(\epsilon_{l},\epsilon_{u})}
{(1+z)} {\rm ln} [2 \epsilon_{l} (1+z)]$ (where
$F(\epsilon_{l},\epsilon_{u})$ is the observed flux between the lower
and upper bandpass limits $\epsilon_{l},\epsilon_{u}$) typically
drops by 3 orders of magnitude, so a $10^{48} {\rm erg \, s^{-1}}$
source only requires a $10^{7} \, M_{\odot}$ black hole. (ii) There is
strong evidence for relativistic beaming in blazars (e.g., through the
observation of superluminal jets (von Montigny et al 1995)). In fact, if blazars were not beamed, we would not be able to see any gamma-rays
from them due to the high pair production opacity at the source;
beaming reduces the luminosity/radius ratio by a factor
$\delta^{p+1}$, allowing photons to escape (Maraschi et al
1992). Here, if $\alpha$ is the
spectral index of the source, then $p= 3+ \alpha$ for a moving sphere in the
Synchrotron Self-Compton (SSC) model, while $p=4 + 2 \alpha$ in the
External Radiation Compton (ERC) model; $\delta=[\gamma(1- \beta {\rm
cos}\theta]^{-1}$ is the relativistic Doppler factor, and $\gamma$ is
the Lorentz factor. If ${\cal
L}$ is the initial intrinsic luminosity of the jet in gamma-ray
emission, beaming
boosts the observed luminosity of the jet to $L= \delta^{p} {\cal
L}$. For $\theta
\sim 0^{\circ}$, then $\delta \sim 2 \gamma$, and the observed
luminosity is amplified by a factor of thousands. The strong
relativistic beaming reduces the fraction of sources which are
visible, since they can only be seen when viewed along the jet
axis. For instance, for $\gamma=6$, and $\alpha=1$, in the SSC model
the observed luminosity is reduced by an order of magnitude from its
maximum if the jet
is pointing $8.5^{\circ}$ from our line of sight, and two orders of
magnitude if the jet is pointing $14.2^{\circ}$ from our line of sight. Note that the black hole
masses derived for a number of low redshift blazars from variability
timescale and transparency arguments lie in the range $10^{7}-10^{8}
M_{\odot}$ (Cheng et al 1999, Hartman et al 1996, Becker \& Kafatos
1995, Romero et al 2000). The luminosity of these blazars is so high
they could be seen at high redshift with GLAST, and black hole masses
of $10^{7}-10^{8} M_{\odot}$ are reasonably abundant at high redshift
in certain models of AGN formation (Haiman \& Loeb 1998). Note that
for a set of blazars of constant intrinsic luminosity ${\cal L}$ and
comoving number density, the redshift distribution of detected sources
in a flux-limited survey flattens considerably and extends to higher
redshifts as the Lorentz factor increases (Dermer \& Gehrels 1995). 

At present, the
modelling of even the low-redshift population of gamma-ray blazars is a matter of considerable
debate. Models which attempt to account for the unresolved gamma ray
background with faint blazars either extrapolate the observed
$\gamma$-ray luminosity function obtained with EGRET (Chiang \&
Mukherjee 1998) or use an assumed conversion between the observed radio loud AGN
luminosity function and the blazar luminosity function
(Stecker \& Salamon 1996). In
this paper, I use a highly simplified model to estimate the
detectability of high-redshift blazars. I assume that the intrinsic
luminosity ${\cal L}$
of the jet in gamma-ray emission (prior to beaming) scales with the luminosity of the
accretion disk ${\cal L}=
f L_{disk}^{\beta}$, where the optical B-band luminosity is taken to
be an accurate reflection of $L_{disk}$ (in particular, assuming the
median quasar spectrum of Elvis et al (1994), a 1 $M_{\odot}$ black
hole shining at the Eddington luminosity has a B-band luminosity of
$5.7 \times 10^{3} {\rm L_{B,\odot}}$). Such a jet-disk correlation is observed in the ratio
of observed radio (i.e., after beaming) to optical luminosities
(Falcke , Malkan \& Biermann 1995). I also assume relativistic beaming with
$L=\delta^{p} {\cal L}$, where $p=3 + \alpha$ (and $\alpha=1$, the
average spectral index observed in the EGRET blazars). The change in
the observed luminosity function due to the effects of relativistic
beaming is given by (Urry \& Padovani 1995):
\begin{equation}
\Phi_{obs}= \int d{\cal L} P(L| {\cal L}) \Phi_{intr} ({\cal L})
\end{equation}
where the probability of observing luminosity $L$ given the intrinsic
luminosity ${\cal L}$ is given by:
\begin{equation}
P(L|{\cal L})= P(\delta) \frac{d \delta}{d L} = \frac{1}{\beta \gamma
\delta} {\cal L}^{1/p} L ^{-(p+1)/p}
\end{equation}
I use the fit to the observed B-band luminosity function $\phi(L_{B},
z)$ from Pei (1995). I adjust the relation ${\cal L}= f L_{disk}^{\beta}$ and the Lorentz factor $\gamma$ (note that since
the distribution of blazar Lorentz factors is unknown, I assume they
all have the same Lorentz factor) to fit the
number of blazars detected by EGRET and their redshift
distribution. I ignore the effects of blazar flaring, which increases
the luminosity by some factor $A$ (where typically $A \sim 5$) some
fraction $\xi$ of the time (where typically $\xi \sim 0.03$), which
results in a second term $ \xi \phi(L/A)$, since these two degrees of
freedom, the normalization (chosen by selecting $\gamma$ and thus the
beaming angle) and luminosity boost (choosen by the combination $L=
\delta^{p} f L_{disk}^{\beta}$), are already present in our model. I find that ${\cal L}_{13}= 3.2 \times 10^{-2} L_{B,13}^{0.9}$ (where
$L_{13}=(L/10^{13} L_{\odot})$ and $\gamma=6$ provides a good fit (see Fig 1, top panel). Note that since $L_{\gamma, 13} < (2
\gamma)^{p} L_{B,13}^{0.9} \sim 600 L_{B,13}^{0.9}$ (corresponding to $\theta=0$), these relations result in a SED
which is in reasonable agreement with the observed SEDs of gamma-ray blazars
(see Fig 1 in Ghisellini et al 1998), where $L_{\gamma}/L_{B} \sim 10-100$
typically, although it can range from $1-1000$. Furthermore, the
Lorentz factor $\gamma \sim 6$ is in reasonable agreement with the
relativistic Doppler factor $\delta < 2 \gamma$ from models which take
into account the SED, time variability, and gamma-ray transparency of
blazars; the derived $\delta \sim 10-20$ (Ghisellini et al 1998). The
somewhat lower Doppler factors I have adopted conservatively
underestimate the number of high-redshift blazars. 

\begin{figure}
\epsscale{1.00}
\plotone{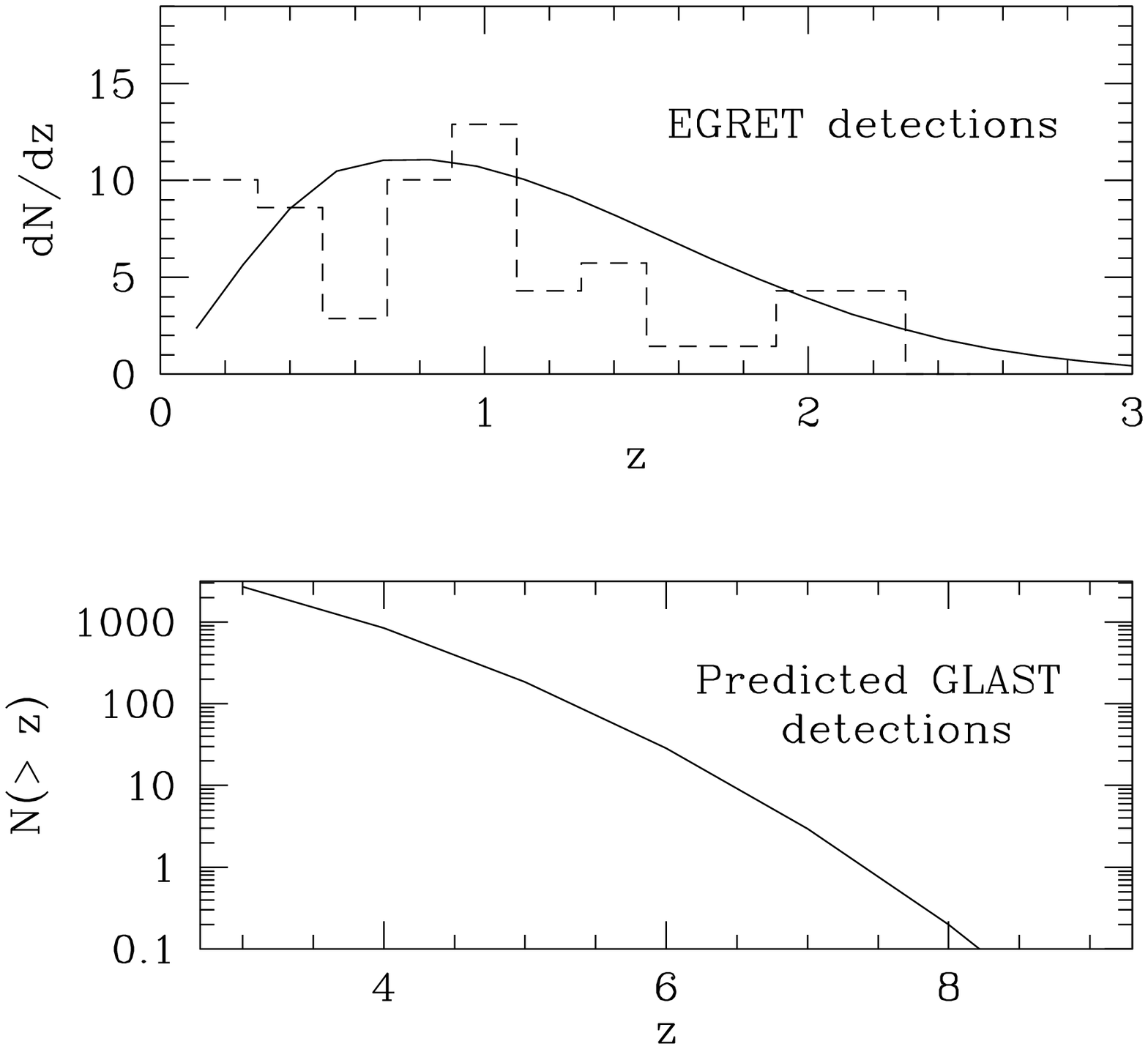}
\caption{Top panel: Best fit model(solid line) for blazar detection
over the entire sky with EGRET detection threshold ${\rm S(E > 100 MeV)} = 2 \times
10^{-7} \, {\rm photons \, s^{-1} cm^{-2}}$, against actual number of
sources detected (dotted line). The model assumes $\gamma=6$ and
$L_{\gamma,13}^{intrinsic}= 3.2 \times 10^{-2} L_{B,13}^{0.9}$. Bottom panel:
predicted number of blazar detections with GLAST detection threshold ${\rm S(E > 100 MeV)} = 2 \times
10^{-9} \, {\rm photons \, s^{-1} cm^{-2}}$ as a function of redshift.
\label{blazar_fig}}
\end{figure}

I then extrapolate this model to high redshift by applying it to the
Press-Schechter based model of Haiman \& Loeb (1998) for quasars. In
this model, which is calibrated to the observed luminosity function of
Pei (1995) at lower redshifts, each halo with $T_{vir} > 10^{4}$K
hosts a black hole with mass $M_{bh} = 10^{-3.2} M_{halo}$ which shines
at the Eddington luminosity for $t_{o} \sim 10^{6}$ years. The result
is shown in the bottom panel of Fig \ref{blazar_fig}, which shows that gamma-ray blazars may
be detected out to $z \sim 7$. By the time GLAST is launched, a large database of quasars with known redshifts will be
available (e.g. from the SLOAN digital sky survey, SDSS (York et al 2000)), and many high-redshift blazars can be selected simply by
identifying their optical counterparts. I emphasize
once again that this highly simplified model is only intended to serve
as a plausibility argument. The main point is that while the Press-Schechter formalism
predicts that massive halos $M_{halo} > 10^{11} M_{\odot}$ expected to
host supermassive black holes of the requisite luminosity become
exponentially rare at high redshift, processes which increase the
luminosity of a lower luminosity population (beaming $L=\delta^{p} {\cal L}$,
flaring $L= A L$) create a power-law tail of bright sources. I have neglected a tail to the distribution of Lorentz
factors, or flaring, which could further flatten the redshift
distribution of detectable sources, increasing the maximum redshift
out to which sources can be seen. Finally, gravitational lensing could
bring otherwise undetectable sources into view, although the low
optical depths for strong lensing (e.g. $\tau \approx 6 \times
10^{-3}$ for $z_{s}=7$, Porciani \& Madau 2000) imply that this should only have a small
impact on number counts. 

\section{Calculating pair production opacity}

The pair-production optical depth for a photon observed at energy $E_{o}$
and emitted from a source at redshift $z_{s}$ is given by (e.g., Madau
\& Phinney 1996):
\begin{equation}
\tau(E_{o},z_{s}) = \int_{0}^{z_{s}} dz \frac{dl}{dz} \int_{-1}^{1}
d(cos\theta) (1-{\rm cos} \theta)  \int_{\epsilon_{th}}^{\infty}
d\epsilon n(\epsilon, z) \sigma (E, \epsilon, \theta)
\label{tau_pp}
\end{equation}
where $E=(1+z)E_{o}$, and $\epsilon_{th}$ is given by the criterion
that pair production can take place, which requires that $E
\epsilon (1 -{\rm cos}  \theta) \ge 2 (m_{e} c^{2})^{2}$. For a given
energy E, the pair production cross-section
$\sigma(E,\epsilon,\theta)$ rises sharply from the threshold energy
$\epsilon_{th}$, reaches a peak of $0.26 \sigma_{T}$ at $\epsilon= 2
\epsilon_{th}$, and finally falls off as $\epsilon^{-1}$ for $\epsilon
\gg \epsilon_{th}$. To calculate the optical depth of the
universe to gamma-ray photons emitted at high redshift, we need to know the
number density of photons as a function of energy and redshift. Before
doing so in detail, I make some simple estimates.

One can perform a simple order of
magnitude estimate to show that the minimal comoving emissivity
required to reionize the universe implies a high pair-production opacity for
$E > E_{th}$, but a low opacity for
$E < E_{th}$,  where we define the threshold energy in the rest frame of
the source for pair production against UV photons at the Lyman edge as
$E_{th} \sim (m_{e} c^{2})^{2}/\epsilon_{L} \approx 18$GeV. This is because the number density of UV photons plummets at wavelengths shortward of
the Lyman limit; the pair production opacity is thus dominated by soft
photons $\epsilon < \epsilon_{L} = 13.6$eV. For the universe
to be reionized, a minimum of 1 ionizing photon per baryon must be
emitted. Thus, the comoving number density of ionizing photons is
$n_{\gamma}(E > E_{L}) > n_{b} (1+ n_{rec}) \sim 10^{-7} (1+ n_{rec})
\ {\rm cm}^{-3}$, where $(1+n_{rec})$ is the average number of times
each baryon recombines during the reionization epoch. The pair
production cross section peaks when $E \epsilon \sim (m_{e}
c^{2})^{2}$, with a value of $\sigma_{pp} \sim 0.26 \sigma_{T}$. Thus, across a Hubble
volume, the optical depth for photons with $E < E_{th}$ is $
\tau \sim n_{\gamma,proper} (E > E_{L}) \sigma_{pp} l_{H} \sim 5 \times
10^{-3} (1+z/10)^{1.5} (1+n_{rec})$, which is undetectably small. What is the pair
production 
opacity for photons with $E > E_{th}$? We can scale the number density
of photons longward of the Lyman limit with respect to the
number of ionizing photons produced. There are 3 factors involved: (i)
$f_{esc}$. Observations of our own Galaxy (Dove, Shull, \& Ferrara 2000,
Bland-Hawthorn \& Maloney 1999) and local starbursts (Leitherer et al
1995) find the escape fraction of ionizing photons from galaxies into
the IGM is of order $f_{esc} \sim 3-6 \%$, and radiative transfer calculations
find that the escape fraction should decrease strongly with redshift
(Woods \& Loeb 1999, Ricotti \& Shull 2000). The
actual comoving number density of ionizing photons produced in starbursts is
$f_{esc}^{-1} n_{b} (1+ n_{rec})$, where $f_{esc} \sim 5 \%$ is the escape fraction
of ionizing photons from the halo. (ii) $f_{break}$. Photoelectric
absorption in stellar atmospheres produces a sharp drop in the flux at
the Lyman edge. For a Salpeter IMF with metallicity
$\sim 10^{-2} Z_{\odot}$, there are $f_{break} \sim 5$ times as many photons emitted
longward of the Lyman limit as there are shortward of it, integrated
over the history of the starburst. (iii) $f_{opacity}$, due to
intergalactic absorption. The energy density $U_{\gamma} \sim \frac{4
\pi}{c} J_{\nu}$ where $J_{\nu} \sim \epsilon_{\nu} \lambda_{mpf}$
where $\lambda_{mfp}$ is the mean free path of an photon of frequency
$\nu$. While the universe is optically thin to photons longward of the
Lyman limit, it is optically thick to ionizing photons, which have a
much shorter mean free path. The energy density of ionizing photons is
lower by a factor $f_{opacity}= \frac{\langle \lambda_{mfp} (E <
E_{L}) \rangle}{\langle \lambda_{mfp} (E > E_{L})\rangle} > 10$ due to
their high absorption rate. The mean free path of ionizing photons
decreases strongly with redshift and for $z >2$ the radiation field is
largely local, due to the large increase in the number of adsorbers
(e.g., Madau, Haardt \& Rees 1999). Thus, we have $\frac{n_{\gamma} (E < E_{L})}{
n_{\gamma} (E > E_{L})}  \sim f_{esc}^{-1} f_{break} f_{opacity} \sim
 10^{3} \left( \frac{f_{esc}}{0.05} \right)^{-1} \left(\frac{f_{break}}{5} \right)\left( \frac{f_{opacity}} {10}
\right) $. This implies that the pair production opacity longward of
$\epsilon_{L}$ over a Hubble
length is much greater, $\tau \sim 5 (1+z/10)^{1.5} (1+n_{rec})$. A similar
estimate may be obtained by normalizing to the observed metallicity of
the IGM at $z \sim 3$, $Z \sim 10^{-2} Z_{\odot}$; this can be shown
to correspond to $\sim 10$ ionizing photons per IGM baryon
(Miralda-Escude \& Rees 1997), or a comoving number density of $\sim
50 n_{b}$ for photons longward of the Lyman break, compared to the
previous estimate of $\sim 100 \left( \frac{f_{esc}}{0.05}
\right)^{-1} \left(\frac{f_{break}}{5} \right) n_{b}$. 

The expectation of a sharp drop in the intensity of the ambient intergalactic
radiation field at the Lyman edge is fairly robust. A
recent composite spectrum of 29 Lyman break galaxies (LBGs) exhibit an
observed flux ratio L(1500)/L(900)=$4.6 \pm 1.0$, which is consistent
with no internal photoelectric absorption, i.e. $f_{esc} \sim 1$ (Steidel, Pettini \& Adelberger
2000). Note that the authors
themselves stress this result should be treated as preliminary; the
result could be due to a large number of uncertainties or selection
effects, among them the fact that these galaxies were selected from
the bluest quartile of LBGs. Even so, if this is typical of all high
redshift galaxies, the observed Lyman break (likely due to absorption
by stellar atmospheres) and expected processing by intergalactic
absorption at high redshift imply a Lyman edge $\frac{n_{\gamma} (E < E_{L})}{
n_{\gamma} (E > E_{L})}  > 50$. Similar considerations of internal and
intergalactic photoelectric absorption apply if
quasars dominate the ionizing radiation field. 

As a photon redshifts, it has to interact with higher energy photons
to pair produce. A photon is able to pair produce until it redshifts
below $E_{th}$, i.e., the universe is optically thin to a photon once
it redshifts below $(1+z_{th}) \sim \frac{E_{th}}{E_{s}} (1+z_{s})$,
where $E_{s}$ is the original energy of the photon at source redshift
$z_{s}$. Thus, there is a fairly well-defined pathlength $\delta l =
\frac{c}{H_{o}} ( (1+z_{s})^{-3/2} - (1+z_{th})^{-3/2})$ over which a
photon may pair produce. This has the fortunate consequence that low
redshift photons do not affect photons emitted near $E_{th}$; thus,
near the threshold energies one is always probing the average radiation field at
redshifts comparable to the source redshift. The flux decrement at a
given frequency measures the average number density of photons redward of
the Lyman break over the associated redshift interval, $n_{\gamma} \sim
\tau/(\Delta l \sigma_{pp})$. Gamma ray absorption measurements thus
provide a fairly clean measurement of radiation fields at high
redshift which are uncomplicated by radiative transfer effects since
(apart from unimportant ${\rm H}_{2}$ opacity effects) the universe is optically thin to
photons redward of the Lyman limit. In particular, photons longward of
the Lyman limit establish a homogeneous, isotropic radiation field
early in the history of the universe. As the pathlengths for pair
production opacity to become significant are typically of order a
Hubble volume, opacity fluctuations due to Poisson fluctuations or
source clustering are insignificant. The measured background radiation field
may be compared directly against the expected background from
measurements of the source luminosity function by direct imaging
of high redshift sources in mid-IR with NGST, where fast photometric
redshifts may be obtained using the Gunn-Peterson break. A comparison of the two
should in principle allow one to check the completeness of a survey at
NGST flux limits.

Let us now use a specific model of high-redshift star formation to
calculation the abundance of UV photons at high redshift. The solution of the cosmological radiative transfer equation yields the mean specific intensity of the radiation background at
the observed frequency $\nu_{o}$, as seen by an observer at redshift
$z_{o}$ as (Peebles 1993):
\begin{equation}
J(\nu_{o},z_{o})= \frac{1}{4 \pi} \int_{z_{o}}^{\infty} dz
\frac{dl}{dz} \frac{(1+z_{o})^{3}}{(1+z)^{3}} \epsilon(\nu,z) e^{-\tau_{{\rm
eff}}(\nu_{o},z_{o},z)}
\label{rad_transfer}
\end{equation}
where $\nu=\nu_{o}(1+z)/(1+z_{o})$, and $\tau_{{\rm
eff}}(\nu_{o},z_{o},z)$ is the effective optical depth of the IGM to
radiation emitted at $z$ and observed at $z_{o}$ at frequency $\nu_{o}$. Redward of the Lyman edge,
radiative transfer is particularly simple as the universe is optically
thin, and only the redshifting of photons is important (there is one caveat to this
statement: the optical depth of the IGM in the Lyman resonance lines
prior to reionization is very large; thus whenever a photon redshifts
into a Ly$\beta$ or higher order Lyman resonance, it is reprocessed
into a Ly$\alpha$ and Balmer or lower order line photon (see Haiman,
Rees \& Loeb 1997)). However, this merely causes a modulation in the
spectrum in the 11.2--13.6 eV range, redistributing photons to the
Ly$\alpha$ and Balmer wavelengths. The large typical energy intervals of
target photons (see Fig 3, top panel) implies that this redistribution
causes the
pair-production opacity to remain the same or increase. I therefore
ignore this complication). The number
density of photons in an energy interval is then given by $\frac{dn}{d
\epsilon} = \frac{4 \pi} {hc} J_{\nu}$. I model the
star formation history $\dot{\Omega}_{*}$ of the high redshift universe with the
semi-analytic models of Haiman \& Loeb (1997), in which a fixed
fraction $f_{star} \sim 1.7-17 \%$ of the gas (normalised to reproduce
the observed IGM metallicity at $z=3$ of $Z=10^{-3}-10^{-2}
Z_{\odot}$) in halos with $T_{vir} > 10^{4}$K (which are able to undergo atomic cooling) fragment to form a starburst lasting for $\sim 10^{7}$ years,
and the halo collapse rate is given
by the Press-Schechter formalism (Sasaki 1994): 
\begin{equation}
\frac{d\dot{N}^{form}}{dM}(M,z)=\frac{1}{D}\frac{dD}{dt}\frac{d n_{PS}}{dM}(M,z)\frac{\delta_{c}^{2}}{\sigma^{2}(M)D^{2}} 
\end{equation}
where $D(z)$ is the growth factor, and $\delta_{c}=1.7$ is the
threshold above which mass fluctuations collapse. In this case the
star formation rate is given by:
\begin{equation}
\dot{\Omega}_{*}(z)= \frac{1}{\rho_{c}(z)} \frac{1}{t_{o}} \frac{\Omega_{b}}{\Omega_{m}} f_{star} \int_{t(z)-t_{o}}^{t(z)} dt
\int_{M(T_{vir}=10^{4} K, z)}^{\infty} dM \frac{d\dot{N}^{form}}{dM}(M,z) M
\end{equation}
An approximate fit to the comoving star formation rate in the
interval $3< z < 10$ can be given by
$\dot{\rho}_{*}={\rm exp}(a0+a1 z+a2 z^{2}) \ {\rm M_{\odot} \, yr^{-1} \, Mpc^{-3} }\,
h^{3}$, where $a0=-0.841,a1=0.395,a2=-0.0295$. The comoving
emissivity is:
\begin{equation}
\epsilon_{\nu}(t)=\rho_{c} \int_{0}^{t} dt^{\prime} F_{\nu} (t -t
^{\prime}) \dot{\Omega}_{*}(t^{\prime})
\end{equation}
where $F_{\nu}(\Delta t)$ is the stellar population spectrum, defined
as the power radiated per unit frequency per unit initial mass by a
generation of stars with age $\Delta t$. I
obtain this spectrum from the Bruzual \& Charlot (1993) code for a
$Z=10^{-2} Z_{\odot}$
population (i.e., extremely low metallicity), assuming a Salpeter IMF
with lower and upper mass cutoffs at 0.1 and 100 $M_{\odot}$, and only
computing the emissivity redward of the Lyman break. I ignore the
effects of dust extinction, which should be negligible at these high
redshifts when the metallicities are very low.  
     
\subsection{Recombination radiation}

Is the number density of recombination line photons such as Ly$\alpha$
sufficiently high to cause significant pair production opacity? It has
been emphasized (e.g., Loeb \& Rybicki 1999) that apart from possible dust attenuation, Ly$\alpha$
photons are not absorbed by the IGM but resonantly scattered until
they redshift out of resonance. Indeed, the Ly$\alpha$ radiation
intensity has been predicted to be particularly strong prior to the
epoch of reionization (Haiman, Rees \& Loeb 1997, Baltz, Gnedin \&
Silk 1998). This is because the optical depth in the Lyman series is
very high, and all the Lyman series lines except Ly$\alpha$ are
absorbed immediately and redistributed. Thus, the Ly$\alpha$ and
Balmer lines are considerably brighter because they receive the energy
of the Lyman series. This has spawned suggestions of detecting the
epoch of reionization by a sharp drop in intensity at the rest-frame Ly$\alpha$
wavelength (Baltz, Gnedin \& Silk 1998, Shaver et al 1999). In
particular, Baltz, Gnedin \& Silk (1998) find that Ly$\alpha$ is about
3 times brighter and H$\alpha$ is about 30 times brighter immediately
prior to the reionization redshift.
 
If this is indeed the case, Ly$\alpha$ photons might contribute
significantly to the pair production opacity. Let us define
$f_{jump}= n_{recomb,Ly\alpha}/n_{tot,Ly\alpha}$, the jump in the
number density of photons longward of the Ly$\alpha$ wavelength. This
is given by the number of photons $n_{recomb,Ly\alpha}$ injected at the Ly$\alpha$
wavelength , over the total number of photons
$n_{tot,Ly\alpha}$, include those redshifting into resonance. Most
studies calculate the number of Ly$\alpha$ recombination photons by summing over the
IGM recombination rate in numerical simulations, by direct estimation
of gas clumping in the simulations. In fact, this underestimates the number of Ly$\alpha$ photons produced: if the escape
fraction of ionizing photons is small, most recombinations occur in the
HII regions of dense halos where star formation takes place, and the primary source of
Ly$\alpha$ photons are the ionising sources themselves. If each ionizing photon is converted into a Ly$\alpha$ photon (plus
lower energy photons) at the source, then
$\dot{n}_{Ly\alpha} \approx \dot{n}_{ion} (1- f_{esc})$, where
$f_{esc}$ is the average escape fraction of ionizing photons from a souce. Thus, inserting $\epsilon
(\nu,z)= E_{Ly\alpha} \dot{n}_{ion} (1- f_{esc}) \delta(\nu - \nu_{Ly\alpha})$ into
equation (\ref{rad_transfer}), I obtain for $\nu < \nu_{Ly \alpha}$,
the solution 
\begin{equation}
J_{\nu}^{Ly\alpha}(z)= \frac{1}{4 \pi} \frac{c}{H_{o}} h_{P} (\frac{\nu}{\nu_{Ly\alpha}})^{-3}
\frac{1}{(\Omega_{m} (1+z_{s})^{3} + \Omega_{\Lambda}
)^{1/2}}\dot{n}_{ion}(z_{L}) (1-f_{esc}) 
\end{equation}  
where ${n}_{ion}(z)$ is the production rate of ionizing photons in
proper coordinates, {\it prior} to attenuation by the ionizing photon
escape fraction. This is a lower bound on $J_{\nu}^{Ly\alpha}(z)$
since it does not include the conversion of photons trapped in higher order
Lyman resonances to Ly$\alpha$ photons. I find that for the adopted stellar spectra, in which
the Lyman edge is typically $f_{break} \sim 5$, then
typically $f_{jump} \sim 1$, which implies that Ly$\alpha$ photons are
comparable to UV continuum photons as a source of opacity. The jump factor $f_{jump}$ may easily be understood as the
ratio of the 
total number of Lyman alpha photons produced (or, the total number of
ionizing photons produced) against the total number of photons in the
$10.2-13.6$eV range. If the break at the Lyman edge is reduced, then
$f_{jump}$ is increased and Ly$\alpha$ photons are more important in
contributing to the pair-production opacity. In particular, for a
power law spectrum with no discontinuity at the Lyman edge (e.g. as
for quasars), Ly$\alpha$ photons are the dominant source of
opacity. Ly$\alpha$ photons are also the dominant source of opacity in scenarios where the universe is reionized by zero
metallicity stars. The higher effective temperature of these Pop III
stars imply that their spectrum is much harder, and significantly fewer
photons are emitted longward of the Lyman break (Tumlinson \& Shull
2000). 
 
Ly$\alpha$ photons could thus an important source of pair production
opacity provided the escape fraction is small, i.e. $(1-f_{esc})
\approx 1$, and dust attenuation while the Ly$\alpha$ photons
resonantly scatter in the IGM is unimportant. If Ly$\alpha$ photons
are the dominant source of pair production opacity, this would be extremely interesting, as it would provide a indirect
census of the ionizing photon emissivity, prior to
attenuation within the host sources. This could prove to be a good
measure of the comoving star formation rate. One way to check if this is the case
would be to directly image sources in NGST in rest frame UV longward of Ly$\alpha$ (sufficiently far away from the Ly$\alpha$
damping wing), as well as in rest frame Balmer
line emission (which should also be directly proportional to the
production rate of ionizing photons, $\dot{n}_{H \alpha} \propto
\dot{n}_{Ly \alpha} \propto \dot{n}_{ion} (1 - f_{esc})$). This will give a sense as to
whether Ly$\alpha$ or UV continuum photons are a greater source of opacity,
provided the trend observed in bright sources extrapolates down
to lower luminosities. 

\subsection{Results} 

\begin{figure}
\epsscale{1.00}
\plotone{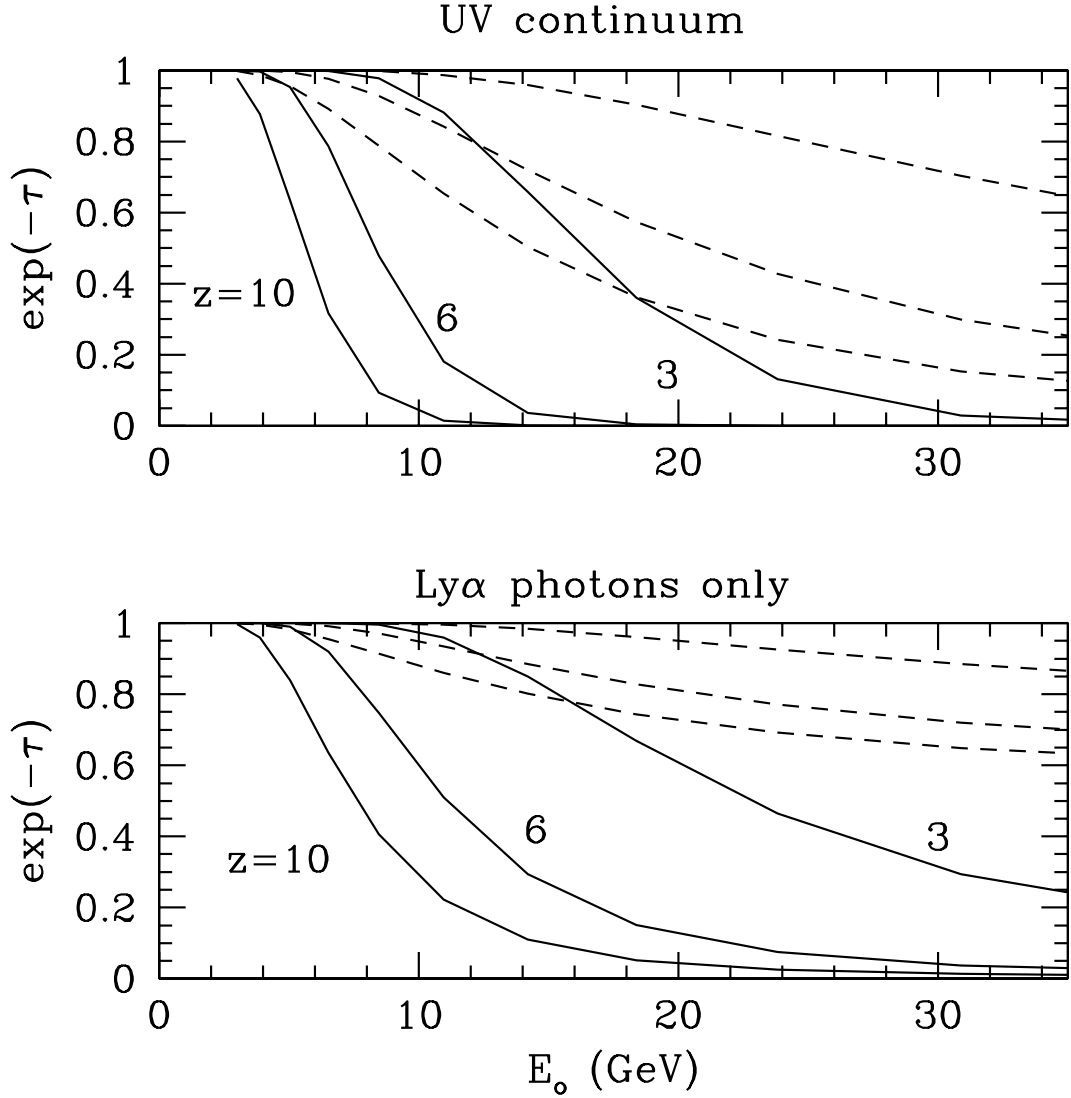}
\figcaption{The predicted attenuation factor as a function of observed photon
energy ${\rm E_{o}}$ for UV continuum opacity (top panel) and Ly$\alpha$
photon opacity only, for source redshifts $z_{s}=3,6,10$. The solid
lines are for a high star formation efficiency
($f_{star}=17\%$), and dashed lines are for a low star formation efficiency
($f_{star}=1.7\%$).
\label{tau_fig}}
\end{figure}

\begin{figure}
\epsscale{1.00}
\plotone{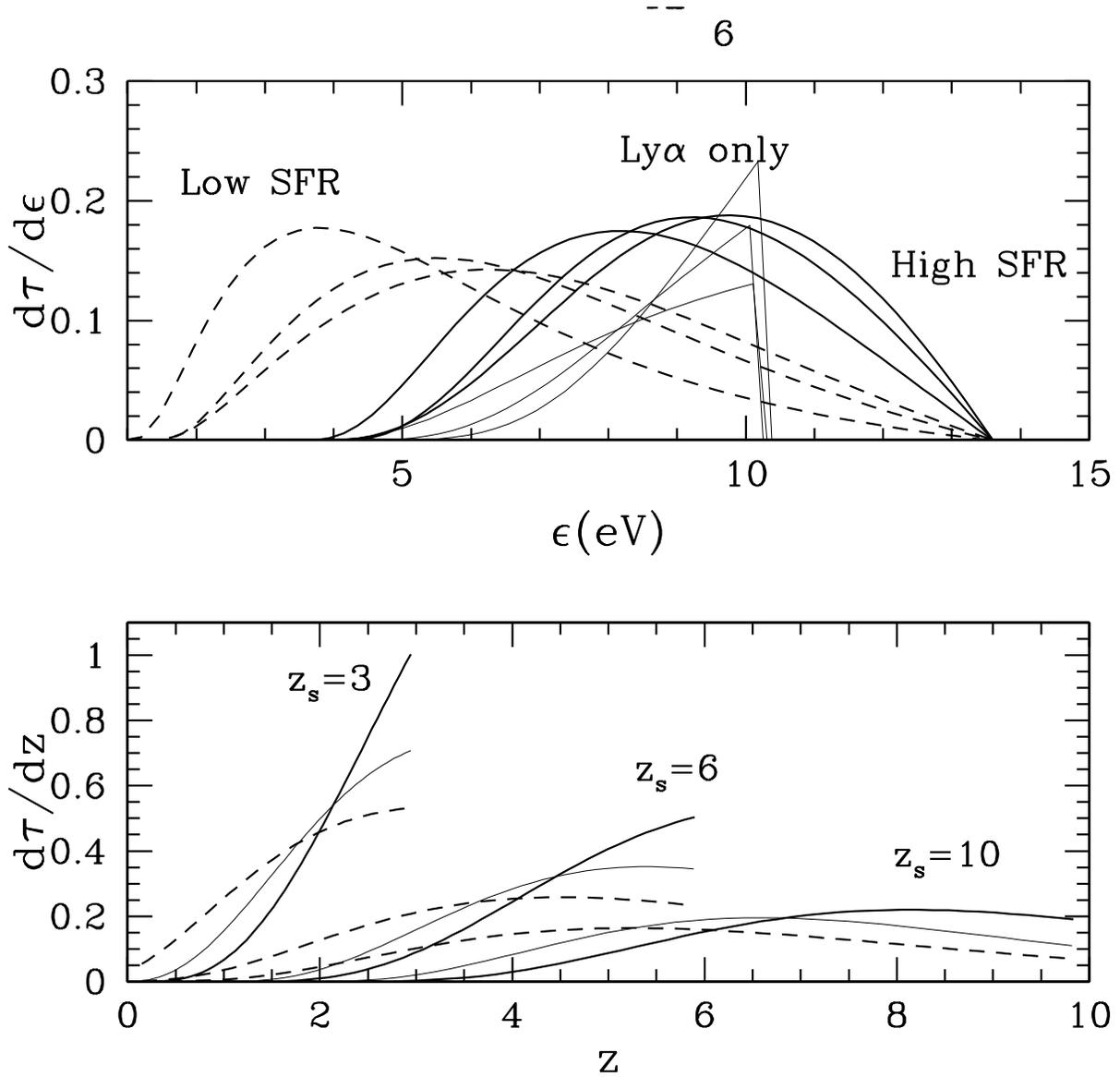}
\caption{The contribution of rest-frame target photon energy interval
(top panel) and redshift interval (bottom panel) to the opacity, all
for gamma-ray photons for which $\tau(E_{o},z_{s})=1$. Figures are for
3 different source redshifts ($z_{s}=3,6,10$) and 3 different UV
emissivity scenarios:high SFR ($f_{star}=17 \%$, dark solid line), low
SFR ($f_{star}=1.7\%$, dashed line), Ly$\alpha$ photon opacity only
(assuming high SFR, thin solid line). Note that contribution of
different energy intervals evolves only weakly with source
redshift ($z_{s}=3,6,10$ from right to left in top figure). As the number density of target photons decreases (low SFR),
the distribution of energy and redshift interval contributions broadens.
\label{dtaudx_fig}}
\end{figure}

Fig \ref{tau_fig} shows the predicted attenuation factor as a
function of observed photon energy, for grazars at redshifts $z_{s}=3,
6,10$, for both high ($f_{star}= 17 \%$) and low ($f_{star}=1.7\%$) star
formation efficiencies. There are several important features to
note. Firstly, the shape of the attenuation curve for Ly$\alpha$ is
similar to that for UV continuum photons. It is not possible to
immediately distinguish between scenarios where Ly$\alpha$ photons and
UV continuum photons are the dominant source of opacity. Secondly, for
lower star formation efficiencies, attenuation both sets in at only at
higher energies and the attenuation curve is significantly
shallower (i.e., it reaches full attenuation after a much longer energy
interval). This is easy to understand. Gamma-ray photons of higher energy have both a longer path-length to travel before they
redshift to $E < E_{th}$, and a higher number density $n(\epsilon_{th} <
\epsilon < \epsilon_{L})$ of photons to pair
produce against, since $\epsilon_{th} \sim (m_{e} c^{2})^{2} /E$ is
lower. Thus, if the
overall number density of UV photons is lower, one must go to higher
energies to achieve the same attenuation. In Figure
\ref{dtaudx_fig}, I show the relative contribution of different
rest-frame target photon energies and redshift intervals for
gamma-ray photons with $\tau(E_{o},z_{s})=1$, for a variety of source
redshifts. The shape of these curves is largely dependent on the
overall number density of target UV photons. For a lower SFR, the energy $E_{o}$ at which
$\tau(E_{o},z_{s})=1$ increases, and the target photon interval and
redshift interval contributing to the resultant opacity broadens. Note that most of the opacity comes from
redshifts comparable to that of the source. Also, the opacity
arises from a fairly narrow target photon energy interval $\sim 3
-13.6$eV; this varies weakly with source redshift. Thus the results
do not depend strongly on the assumed spectral
slope of the UV sources. I also obtained by direct computation the contribution to the opacity due
to ionizing photons shortward of the Lyman edge. Depending on one's assumptions for
$f_{esc},f_{break},f_{opacity}$, it is smaller by 2--4 orders of
magnitude, and is completely negligible.   

\begin{figure}
\epsscale{1.00}
\plotone{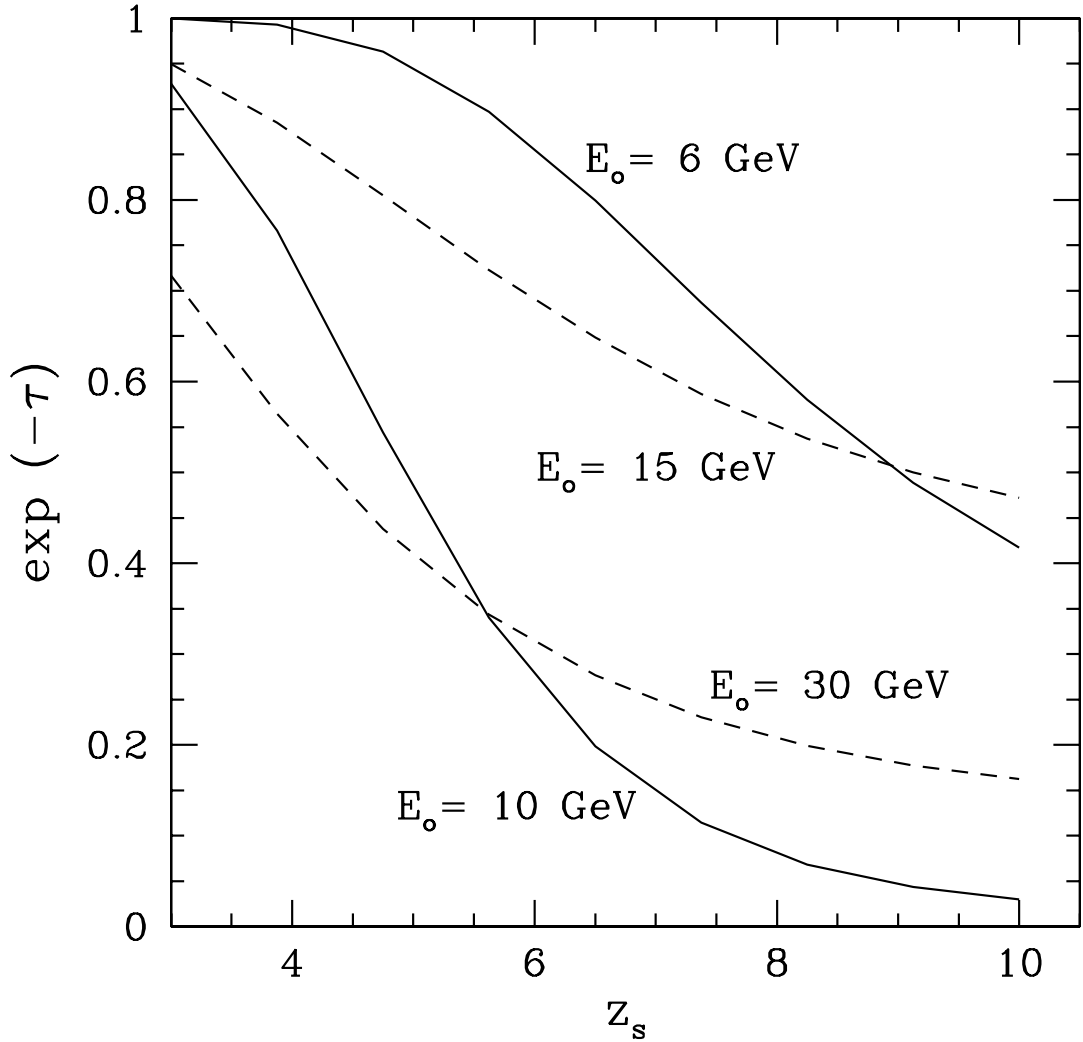}
\caption{The attenuation factor as a function of blazar redshift for a
given observed photon energy. Solid curves are for the high star
formation efficiency case, dashed curves are for the low star
formation efficiency case. Any increase in the attenuation between
$z_{1}$ and $z_{2}$ is solely due to UV radiation in the range $z_{1} < z
< z_{2}$. Note that if the star formation efficiency is low,
the degree of attenuation changes much more slowly with redshift.
\label{tau_evolve}}
\end{figure}

\begin{figure}
\epsscale{1.00}
\plotone{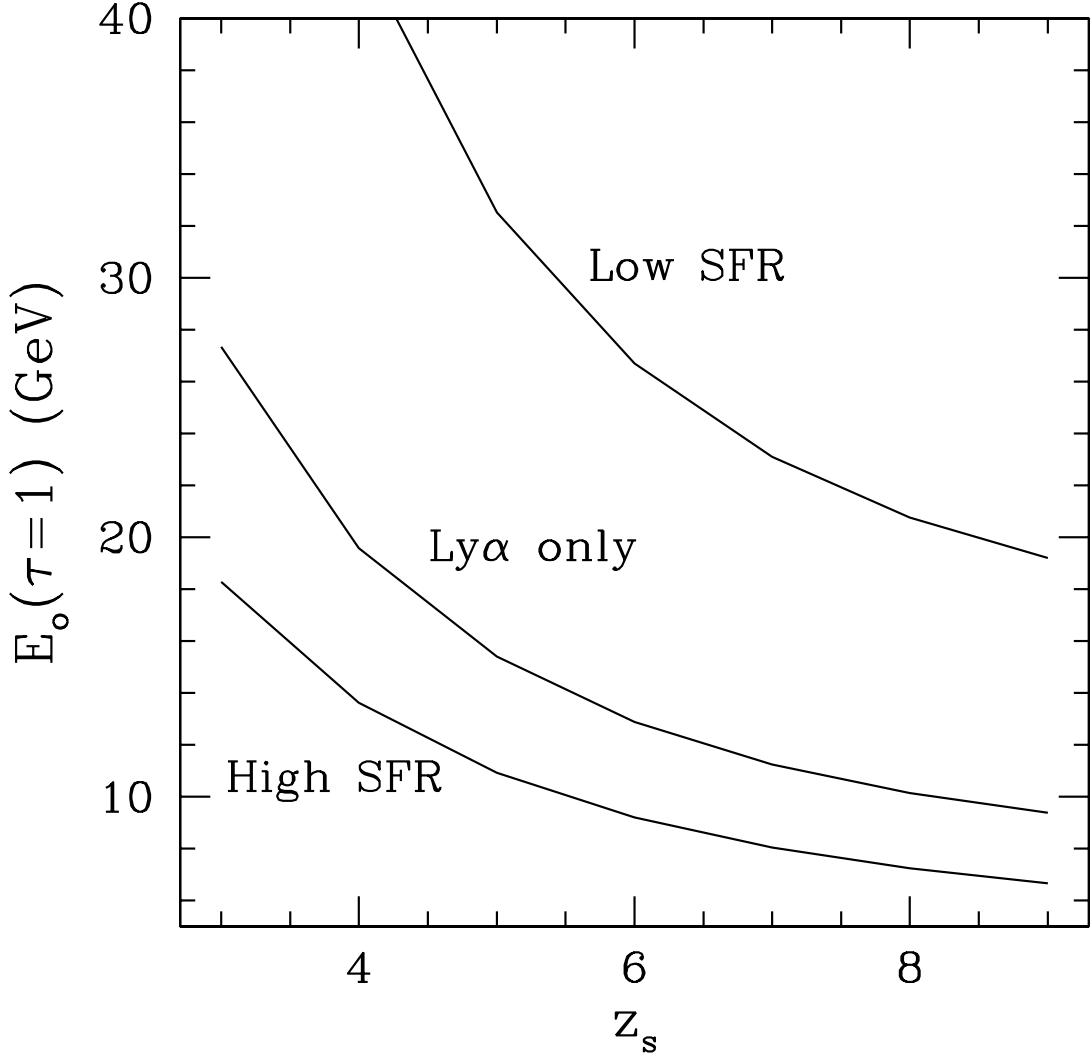}
\caption{The variation of the observed energy $E_{o}$ at which $\tau=1$ with
source redshift, for a high star formation efficiency
($f_{star}=17\%$), a low star formation efficiency ($f_{star}=1.7\%$),
and a high star formation efficiency model where Ly$\alpha$ photons
provide the main source of opacity. Note how the curve flattens at
high redshift due to the reduced opacities at high redshift. If the
pair production opacity at high redshift is negligible, the curve
would be completely flat.
\label{tauone_redshift}}
\end{figure}

An important technique for isolating the contribution of high-redshift
radiation fields would be to use measurements of {\it differential}
absorption between blazars at different redshift. Consider two blazars
with at redshifts $z_{1}$ and $z_{2}$, with $z_{2} > z_{1}$. At a
given observed energy $E_{o}$, photons from each blazar will encounter
identical optical depths for $z< z_{1}$. Thus, any additional opacity seem
in the spectrum of the blazar at $z_{2}$ must arise from the redshift
interval $z_{1} < z < z_{2}$ alone, $\Delta \tau (E_{o}) = \int_{z_{1}
}^{z_{2}} dz \frac{dl}{dz} \int_{-1}^{1}
d(cos\theta) (1-{\rm cos} \theta)  \int_{\epsilon_{th}}^{\infty}
d\epsilon n(\epsilon, z) \sigma (E, \epsilon, \theta)$. Note that this
identification is independent of any uncertainties in the
spectral slope or redshift evolution of the UV background. As argued above, radiation field intensity
fluctuations are unimportant since the pair-production opacity arises
on scales of order a Hubble length. Thus, in
Figure (\ref{tau_fig}), any difference between the $z_{s}=3,6, 10$
curves is solely due to radiation fields between $3 < z < 6$ and $6 < z <
10$; if star formation ceases for $z > 3$, the curves would lie on top
of one another. If there is little difference
between two absorption curves from sources at $z_{1}$ and $z_{2}$, one
can get an upper bound on the UV emissivity in the range $z_{1}
< z < z_{2}$; likewise, if two curves are so widely separated that
meaningful measurements of $\Delta \tau (E_{o})$ cannot be obtained
(in particular, if absorption saturates in one of the curves),
one can get a lower bound on the UV emissivity in the range $z_{1}
< z < z_{2}$. In Figure (\ref{tau_evolve}), I show how the attenuation
at a fixed observed photon energy $E_{o}$ is expected to rise with
redshift, due to the opacity provided by high redshift UV fields. For
lower star formation efficiency, the attenuation rises more slowly; if
there is no star formation at high redshift the curve would be
flat. In Figure (\ref{tauone_redshift}) I show how the energy $E_{o}$
for which $\tau(E_{o},z_{s})=1$ is expected to fall with increasing redshift; due
to the increased opacity provided by high redshift photons,
attenuation sets in at lower energies. Again, if
there were no star formation at high redshift, this curve would show no evolution.   

In fact, differential absorption measurements provide a powerful
cross-check on the technique: if $\tau(E_{o},z_{2}) <
\tau(E_{o},z_{1})$, then the assumed
unabsorbed blazar spectrum must be evolving with redshift, due to a
change in internal absorption, or underlying spectral index. Note that
if star formation rates are high, the attenuation occurs rapidly over a small energy interval,
and uncertainties due to the extrapolation from the unabsorption
portion of the spectrum is small; for lower star formation rates the
attenuation occurs over a larger energy interval, and uncertainties
due to extrapolation are greater.  

\section{Conclusions}

In this paper, I have suggested that if blazars can be detected at high
redshift, detection of gamma-ray absorption due to pair production
against high-redshift UV photons will provide a valuable probe of
high-redshift UV radiation fields. This is because the sharp Lyman
edge in the intergalactic radiation fields implies that gamma-ray
photons have only a limited redshift interval in which to
pair-produce. As they redshift to lower energies, they require photons with $\epsilon > 13.6$eV to
pair-produce, so the
universe becomes optically thin. The shape of the attenuation curve is
primarily sensitive to the
overall number density of photons longward of the Lyman edge at high
redshifts: the higher this number density, the lower the gamma-ray
photon energy at which pair-production opacity sets in. This makes it a
useful test of the overall level of star formation and ambient UV
radiation fields present at high redshift. Ly$\alpha$ photons
provide an important contribution to this pair production opacity, and
indeed may be the dominant source of opacity if sources with a
relatively lower fluxes longward of the Lyman edge (such as quasars or low metallicity stars) are
abundant. Finally, measurements
of differential absorption between blazars at the same observed
energies will allow us to cleanly isolate the increase in opacity due to
radiation fields at high redshift.    

There are two large uncertainties. The first is whether GLAST will be
able to see high-redshift blazars at all. However,
in the unified model of AGN, the scarcity of ultra-luminous blazars is a geometrical
effect (due to relativistic beaming) rather than a requirement of
extremely large black hole masses. In fact, the luminosity boost
provided by beaming reduces the black hole mass by several orders of
magnitude below that demanded by the Eddington limit. So it is at least
plausible that high redshift blazars will be detectable. The second
uncertainty is whether absorption seen in a blazar will be internal,
rather than due to pair production against photons in the
IGM. However, at GeV energies we have some physical understanding of
the observed EGRET spectra (e.g., Ghisellini et al 1998); opacity to
gamma-ray photons due to internal radiation fields can be constrained
by time variability arguments and other constraints. Furthermore, GLAST should assemble an extremely
large catalog ($>$ few thousand) of low redshift blazars, whose spectra
can be studied in detail (and the contribution to opacity due to low
redshift star formation can be quantified by other means); provided
blazar properties do not evolve too strongly with redshift, we should
have a firm handle on the intrinsic unabsorbed blazar spectrum. 

\section{Acknowledgements}

I thank David Spergel for his advice and encouragement, Youjun Lu
for helpful conversations on blazar properties, Bruce Draine for helpful comments, and Bart Pindor for
technical assistance. I thank the Institute of Theoretical Physics,
Santa Barbara for its hospitality during the completion of this
work. This work is supported by the NASA ATP grant NAG5-7154, and by the National Science Foundation, grant number PHY94-07194.   



\end{document}